\documentclass[12pt,preprint]{aastex}

\usepackage{graphicx}

\newcommand{\trans}[2]{J={#1}$\rightarrow${#2}}
\newcommand{\HISA}{{\sc HI}SA}

\shorttitle{CO in  \HISA\ Clouds}
\shortauthors{Klaassen et al.}

\begin{document}

\title{CO in HI Self-Absorbed Clouds in Perseus}

\author{P. D. Klaassen\altaffilmark{1} \altaffilmark{2}, R. Plume\altaffilmark{1},
S. J. Gibson\altaffilmark{1}, A.R. Taylor\altaffilmark{1}, C. M.
Brunt\altaffilmark{3}}

\altaffiltext{1}{Department of Physics and Astronomy, University
of Calgary, 2500 University Dr. NW, Calgary, AB, T2N 1N4, Canada}
\altaffiltext{2}{Currently at Department of Physics \& Astronomy, McMaster University, 1280 Main St. W
Hamilton, ON, L8S 4M1, Canada}
\altaffiltext{3}{Department of Astronomy,  University of Massachussetts, Amherst, Lederle Graduate Research Tower, MA 01003}

\begin{abstract}

We have observed  $^{12}$CO $J = 2\rightarrow1$ and $J = 1\rightarrow0$, and 
$^{13}$CO  $J = 1\rightarrow0$ emission in  two regions of HI Self-Absorption (\HISA) in Perseus: a small, isolated \HISA\ feature called the {\it globule} and a more extended \HISA\ cloud called the {\it complex}. Using both Large Velocity Gradient  and Monte Carlo radiative transfer codes we found that, in the globule, N($^{12}$CO)  $< 6.0 \times 10^{15}$ cm$^{-2}$ which, using PDR models, implies
that  N(H$_2$) $< 9.9 \times 10^{20}$ cm$^{-2}$.
In the complex we found that the H$_2$ column densities ranged from  $1.2 - 2.2 \times 10^{21}$ cm$^{-2}$.
By comparing the \HISA\ and CO observations we are able to constrain the physical conditions 
and atomic gas fraction ($f$).  In the globule,  8 K $< T_{spin} < 22$ K and $0.02 < f < 0.2$ depending on whether the (unknown) gas density is $10^2$, $10^3$, or $10^4$ cm$^{-3}$.  In the complex, 12 K $< T_{spin} < 24$ K, $0.02 < f < 0.05$, and we were also able to constrain the gas density ($100 < n < 1200$ cm$^{-3}$).  These results imply that the gas in the \HISA\ clouds is colder and denser than that
usually associated with the atomic ISM and, indeed, is similar to that seen in molecular clouds.
The small atomic gas fractions also imply that there is a significant molecular component in these \HISA\ clouds, even when little or no $^{12}$CO is detected. The level of $^{12}$CO detected and the visual extinction due to dust is consistent with the idea that these \HISA\ clouds are undergoing a transition from the atomic to molecular phase.  

\end{abstract}


\section{Introduction}

The study of cold (T $< 100$ K) atomic gas, a major component of the 
interstellar medium in the Galaxy, is a difficult problem.
While {\sc HI} emission lines can be used to easily map the distribution of the 
atomic gas, it is often difficult to separate the warm and cold components.  
On the other hand, direct observations of cold {\sc HI} gas can be obtained from 
{\sc HI} lines seen in {\it absorption} against warm, background {\sc HI} emission 
(called {\sc HI} self-absorption or \HISA ).
Studies of HISA in molecular clouds have shown
that the HISA is often well mixed with molecular gas (Jackson et al. 2002,
Li \& Goldsmith 2003).

Using data from the Canadian  
Galactic Plane Survey (CGPS; Taylor 1999, Taylor et al. 2003), Gibson et al. 
(2000; 2005a; 2005b) have revealed \HISA\ in unprecedented detail over a 
square degree in Perseus.  
In  Gibson et al. (2000; hereafter G2000) two regions in the Perseus arm were found  
to be of particular interest.  The first, labeled the $globule$, is a compact (unresolved in  
the 1$'$  main beam of the CGPS), dark (absorption line center contrast  
$>$ 42 K; Gibson et al. 2005b)  region with narrow ($\Delta V_{FWHM} = 2.5$ km  
s$^{-1}$) \HISA\ substructure. IRAS 60 $\mu$m dust continuum emission was  
observed
in the globule, but there was  
no detected $^{12}$CO \trans{1}{0} emission (Heyer et al. 1998). The second,  
labeled the $complex$, is part of a larger {\sc HI} region.  It also shows a  
deep \HISA\ feature (line center contrast $\sim$ 28 K; Gibson et al. 2005b) and 60  
$\mu$m emission.
However, unlike  
the globule, it has detected $^{12}$CO \trans{1}{0} emission (T$_R^*\approx$  
1.12 K; Heyer et al. 1998).  Figure \ref{fig:region} shows the general  
region of study, with both the globule and complex labeled.
Unfortunately, while  \HISA\  lines can be readily identified throughout the Galaxy, 
 it is difficult to extract physical parameters (density, temperature, column density etc.) from the observations.

Using a new technique, G2000 were able to set limits on mass, temperature, optical  
depth, density and column density in the globule and the complex.  However, the  
range of these limits was quite large, 
mainly due to the unknown molecular gas fraction in the two  
regions.  In the globule, no $^{12}$CO \trans{1}{0} was detected to the  
limiting sensitivity of the Outer Galaxy Survey of Heyer et al. (1998) (hereafter OGS). 
In the complex, $^{12}$CO \trans{1}{0} was  
detected but, with only one molecular transition, it is difficult to  
determine the total $^{12}$CO column density.  Therefore, in this paper, we  
present observations of $^{12}$CO \trans{1}{0} and \trans{2}{1} in both the  
globule and the complex in order to determine the molecular gas  
content.  Our  goals are to complement the \HISA\ observations  
presented in G2000, and to better constrain the physical properties of  
the gas in these two regions.

\section{Observations}
\label{obs}

Our observations of the \trans{1}{0} transition of $^{12}$CO ($\nu$ = 115.3  
GHz) and  $^{13}$CO ($\nu$ = 110.2 GHz) were obtained at the Five  
Colleges Radio Astronomy Observatory (FCRAO) in 2000 February and  
May.  These $^{12}$CO data were obtained in order to either detect the  \trans{1}{0}  
or, at least, achieve a better sensitivity than the OGS  
data.  The $^{13}$CO data were obtained to help determine whether the multiple spectral features
seen in some of the $^{12}$CO observations were due to multiple line-of-sight clouds or self-absorption in the $^{12}$CO.
The FCRAO  14 m telescope has a  
full width half maximum (FWHM) beamsize of 45.5$''$ at 115 GHz, and  
main beam efficiency ($\eta_{mb}$) of 42\%.  The velocity resolution (channel width) of  
these observations is 0.252 km s$^{-1}$. 

The \trans{2}{1} transition of  
$^{12}$CO ($\nu$ = 230.6 GHz) was observed at the Caltech Submillimeter  
Observatory (CSO) in 2001 September.  The CSO is a 10.4 m telescope on  
Mauna Kea, Hawaii, with a FWHM beamsize of 32$''$, and $\eta_{mb}$ of  
69\% (at 230 GHz).  The velocity resolution of these observations was
0.063 km s$^{-1}$. 

\begin{figure}
\begin{center}
\plotone{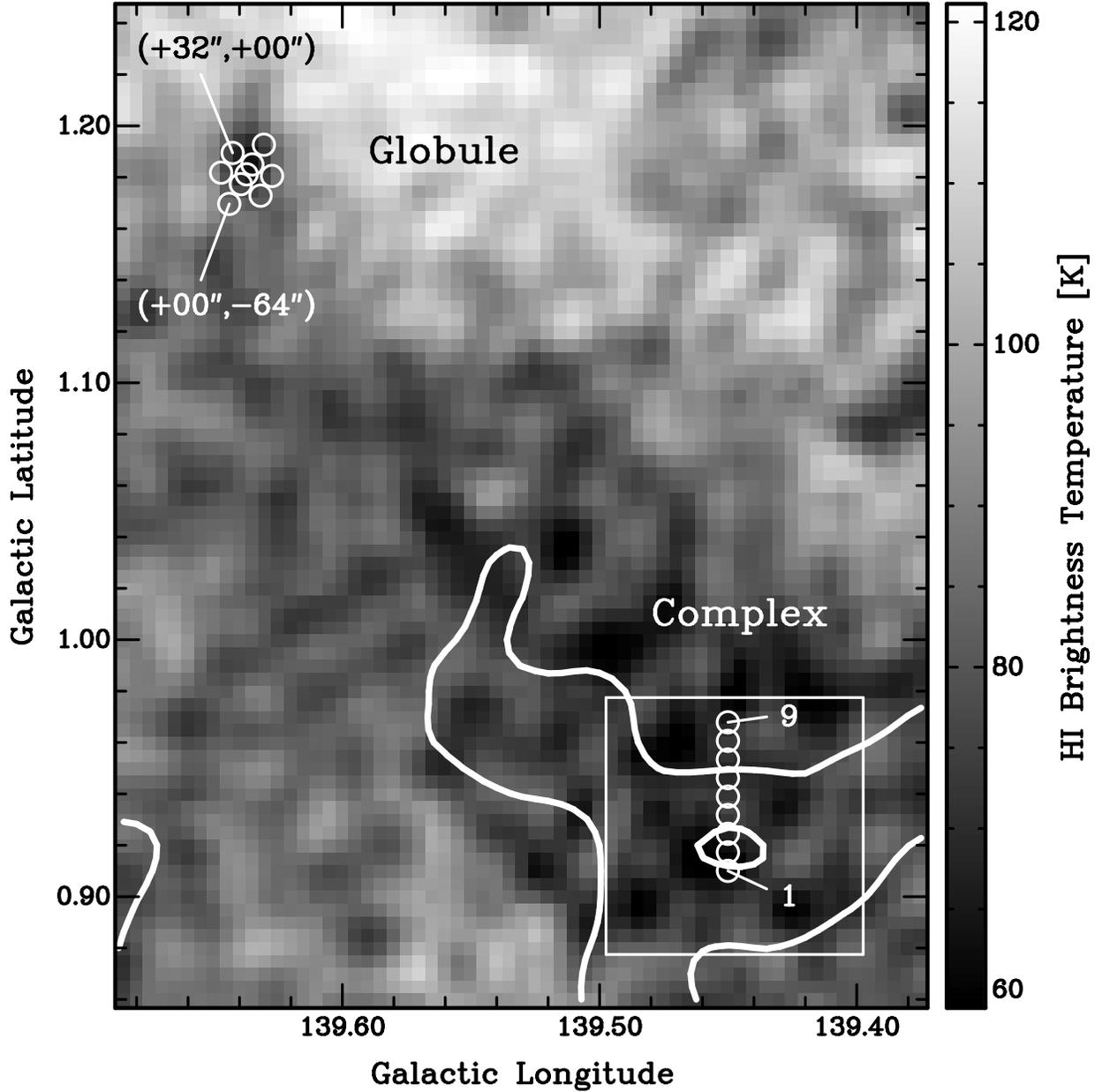}
\caption{\textbf{Region of study:} The complex and globule, as  
defined by Gibson et al. (2000).  The grayscale background is the HI brightness and the contours
denote the $^{12}$CO J = $1 \rightarrow 0$ brightness at 1 and 3 K (from Heyer et al. 1998).  The positions  
which were mapped are labeled in the inset image.  The central coordinates of the
$globule$ are: $\ell \sim 139.635^{\circ}$, $b \sim +1.185^{\circ}$ or $\alpha(J2000) = 3^h09^m24.0^s$,  $\delta(J2000) = +59^o30'22.5''$.  The central coordinates of the $complex$ (labeled as Position 1) 
are $\ell \sim  139.45^{\circ}$, $b \sim
+0.91^{\circ}$ or $\alpha(J2000) = 3^h07^m3.5^s$,  $\delta(J2000) = +59^o21'38.5''$.
}
\label{fig:region}
\end{center}
\end{figure}

In the globule, we observed a single point at the central position in  
the \trans{1}{0} transition and an 8 point map with full CSO (32$''$)  
beam spacings in the \trans{2}{1} transition.    A ninth position 
at a half beam spacing ($\Delta\alpha = 0, \Delta\delta = -16''$) was  
also observed at the CSO. In the  
complex, we observed a 9 point strip in both $^{12}$CO transitions,  as  
well as the \trans{1}{0} transition of $^{13}$CO.  The strip starts at  
the central position (labeled as Position 1 in Figure \ref{fig:region}), 
where $^{12}$CO \trans{1}{0} emission had been detected  
in the OGS, and moves outwards at constant Galactic longitude but  
increasing Galactic latitude (from $b$ = +0.91$^{\circ}$ to  
$+0.97^{\circ}$ in steps of 25$''$) to where $^{12}$CO \trans{1}{0} had not  
been previously detected. The spacing of the complex observations were  
approximately equal to
FCRAO half beam spacings. The  observed positions in both the globule  
and the complex are shown in Figure \ref{fig:region}.

\section{Results}
\label{results}

In Sections \ref{spectra} and \ref{models} that follow, we first provide details of the observed spectra and then show how the results are used to obtain the molecular gas column densities.

\subsection{Description of the Spectra}
\label{spectra}

\begin{figure}
\begin{center}

\includegraphics[scale=0.85, angle=-90]{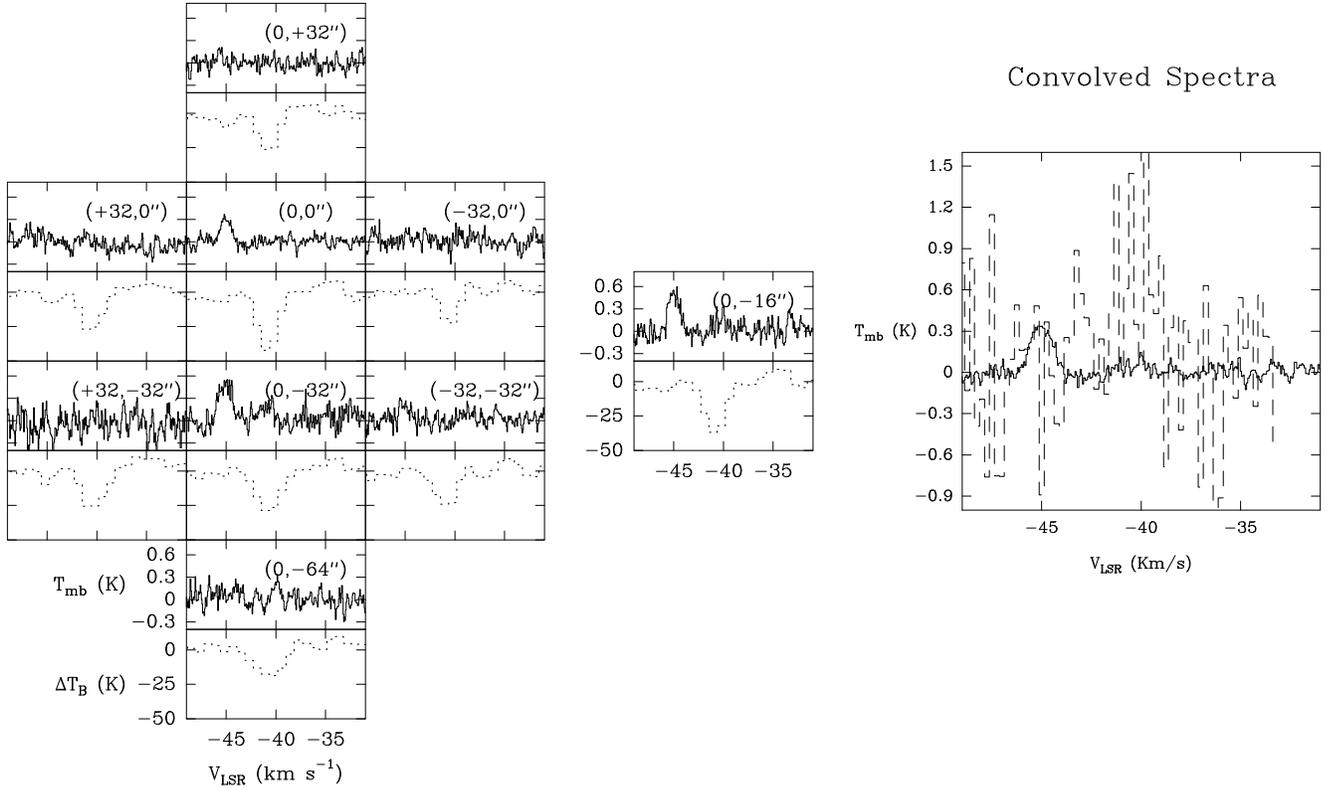}
\caption{ \textbf{\textit{Left  
Panel:}} \trans{2}{1} map of the globule taken at the CSO with a 32$''$  
beam. Note that the half beam spacing observation has been offset (to  
the right) for clarity. The top portion of each position shows the  
\trans{2}{1} transition of $^{12}$CO (solid line), while the bottom portion shows the 
\HISA\ at that position (dotted line). 
\textbf{\textit{Middle Panel:}} Same as above but shows the single half-beam spaced observation in 
the globule (see text).
\textbf{\textit{Right Panel:}} Superposition of  the
$^{12}$CO  \trans{2}{1} CSO observations convolved to the FCRAO 45$''$ beam (solid lines) and \trans{1}{0}
FCRAO observations (dotted lines) of the globule.
}
\label{fig:observations}
\end{center}
\end{figure}

Figure \ref{fig:observations} shows a postage stamp map of the  
$^{12}$CO \trans{2}{1} transition (solid line) and the \HISA\ (dotted  
line) in the globule, with the single half-beam spaced ($\Delta\delta$  
= -16$''$) observation shown in the middle panel for clarity.  Gaussian profiles
were fit to the spectra, the parameters of which are listed in Table \ref{tab:glob} under the heading 
{\it 32$''$ Resolution}.  Positions with no detected emission are listed with their $1\sigma$ rms
noise limits. 

To  facilitate comparison of the \trans{1}{0} and \trans{2}{1} transitions,  
the \trans{2}{1} observations were convolved to match the 45$''$ beam  
of the FCRAO.  The far right panel in Figure \ref{fig:observations}  
shows the convolved $^{12}$CO \trans{2}{1} spectrum (solid line) along with  
the FCRAO \trans{1}{0} spectrum (dashed line) at the central position of the globule.  The $^{12}$CO \trans{2}{1}  
emission is clearly visible in the convolved spectrum, but below the  
peak-to-peak noise level of the \trans{1}{0} observations.  This shows  
that molecular gas is present, but that even our deeper $^{12}$CO \trans{1}{0}  
observations did not reach the sensitivity limit required to detect it.  
  A Gaussian profile was  fit to the \trans{2}{1} transition, the  
parameters of which are listed in Table \ref{tab:glob} under the heading 
{\it Convolved to 45$''$ resolution}. Since there  
was no signal detected in the \trans{1}{0} transition, the 1$\sigma$  
rms noise was used as an upper limit on signal strength.

Figure \ref{fig:observations} clearly shows that, while the \HISA\  
component of the globule lies at $V_{LSR}$ = -41 km s$^{-1}$, the $^{12}$CO  
emission is centered at $V_{LSR}$ = -45 km s$^{-1}$. This suggests that  
these two emission lines are tracing separate clouds  along the line of sight. This 
suggestion is supported by recent observations of the region with  an  
increased signal to noise ratio in the $^{12}$CO \trans{1}{0}  transition  
(Brunt, 2005) which reveal a number of weak $^{12}$CO knots in the  
vicinity of the globule at a $V_{LSR}$ of -45 km s$^{-1}$.  To draw comparisons with the  
atomic globule at $V_{LSR}\approx$ -41 km s$^{-1}$ we use the 1  
$\sigma$ rms noise limit for both the \trans{1}{0} and \trans{2}{1}  
transitions ($T_{mb}$ = 0.58  and 0.06 K respectively) as an upper limit  
on line strength.  As an approximation to a line width, we use the  
FWHM of the \HISA\  lines ($\Delta$V$_{FWHM}$ = 2.5 km s$^{-1}$).  This is a valid assumption if the atomic and molecular gas  
components are mixed and the line widths are dominated by turbulence.  This assumption is supported
by the similarities between the \HISA\ and $^{12}$CO line widths seen in the complex.   These line  
parameters are used  in Section \ref{glob} to constrain the molecular gas content of the  
globule.

\begin{center}
\begin{figure}
\includegraphics[scale=0.85, angle=-90]{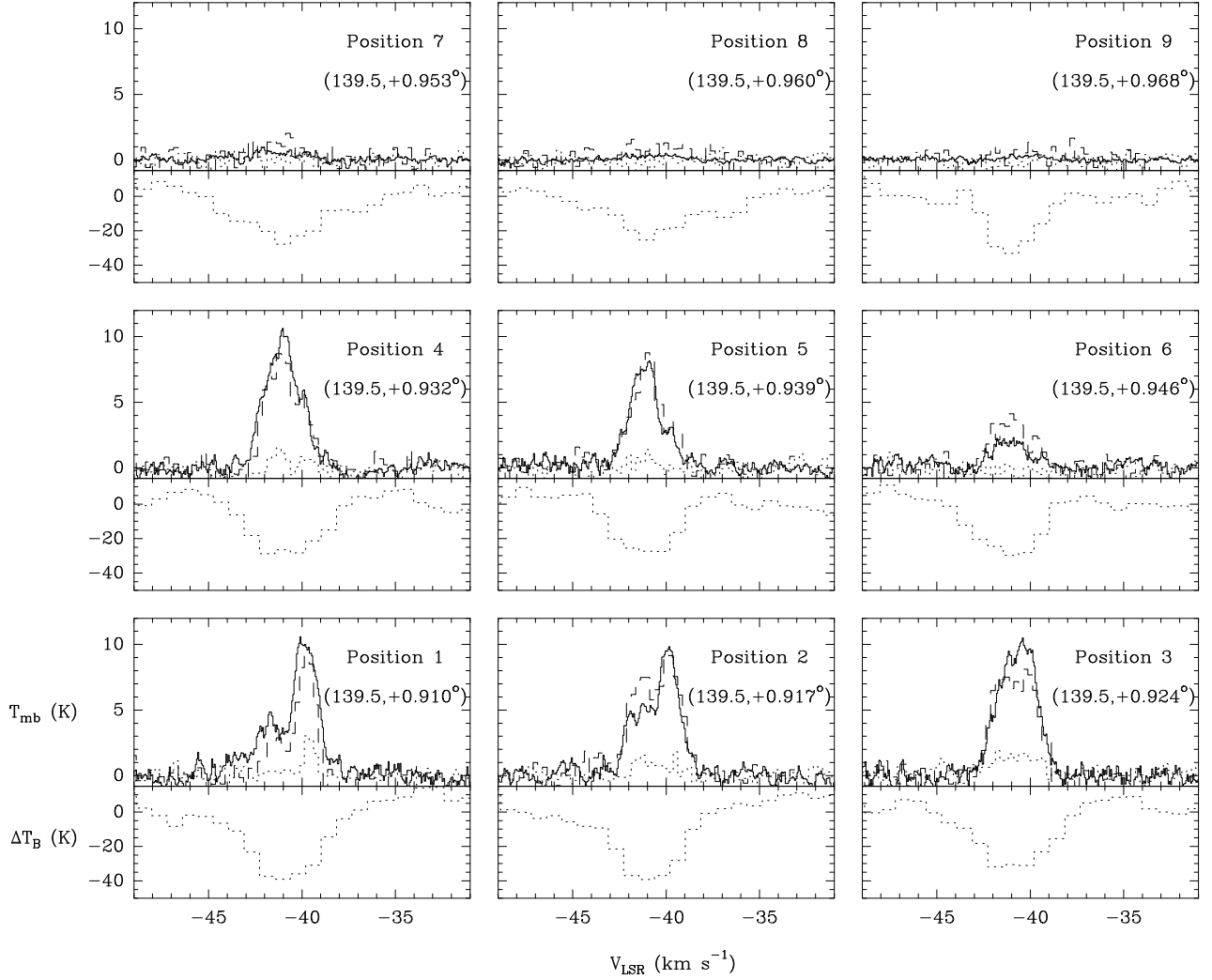}
\caption{Individual spectra for each position in the complex (labeled 1  
through 9). The bottom panel of each position shows the \HISA\,  
profile, while the top panel shows the $^{12}$CO J=2-1 (solid), J=1-0  
(dashed) and $^{13}$CO J=1-0 (dotted) profiles at the position.  The Galactic longitude
and latitude are given in the parentheses in each panel (in degrees).  }
\label{fig:strip_overlay}
\end{figure}
\end{center}

The bottom half of each panel in Figure \ref{fig:strip_overlay} shows  
the \HISA\ as a function of position in the complex. The top halves  
show the $^{12}$CO \trans{2}{1} (solid line), $^{12}$CO \trans{1}{0}  
(dashed line), and $^{13}$CO \trans{1}{0} (dotted line) at the same  
positions. Note that Figure \ref{fig:strip_overlay} is a  
single strip of observations (of increasing Galactic latitude) and is  
only presented as what appears to be a nine point map for readability.  
Figure \ref{fig:strip_overlay} shows interesting changes in  
the spectral line profiles progressing from the cloud core (Position 1)  
to the edge (Position 9). 
 The first  
two positions show emission peaks at $\sim$ -40 km s$^{-1}$ with  
blueshifted shoulders (or secondary peaks) at $\sim$ -41 km s $^{-1}$.  The third 
position is singly peaked, while the fourth through sixth positions have  
 emission peaks at -41 km s$^{-1}$ with {\it redshifted}  
shoulders (or secondary peaks) at -40 km s$^{-1}$.  The emission drops off rapidly after  
Position 6, with Positions 7 through 9 showing no significant signal.  
$^{13}$CO and $^{12}$CO observations show similar line profiles, which  
is suggestive of emission from multiple clouds (as opposed to self  
absorption).  We will test this assumption in the next section.   
Gaussian profiles were fit to the spectra in the complex and are listed  
in Table \ref{tab:com}.  Note that Positions 1 and 2 were fit with  
2 separate Gaussians.

\begin{figure}
\begin{center}
\includegraphics[scale=0.75, angle=-90]{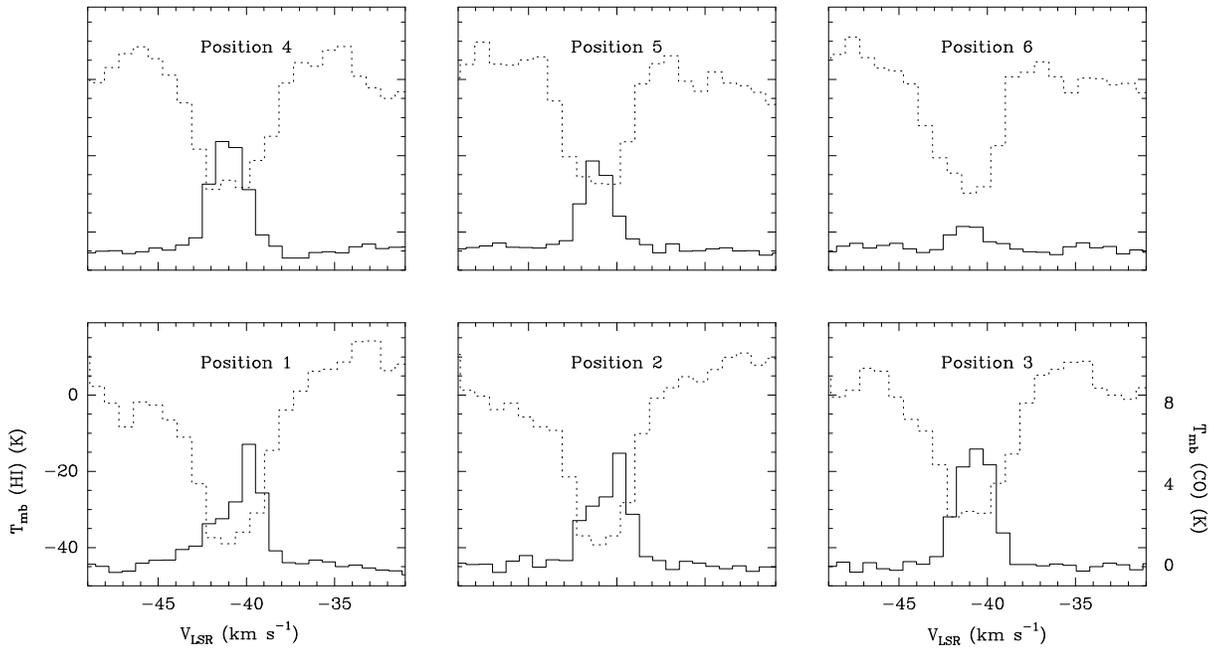}
\caption{$^{12}$CO \trans{2}{1} Observations (solid line) in the  
complex smoothed to the same velocity resolution as the \HISA observations (dashed  
lines).  }
\label{fig:smoothCO}
\end{center}
\end{figure}

Figure \ref{fig:smoothCO} shows the result of smoothing the $^{12}$CO  
\trans{2}{1} in the complex to the velocity resolution of the HI  
observations (solid lines of $^{12}$CO \trans{2}{1} are overlayed on the  
dotted \HISA\ profiles). This spectral resampling blends the multiple $^{12}$CO  
components (seen best in Positions 1 and 2) into one component. While  
there is general agreement between the \HISA\ and the smoothed $^{12}$CO  
emission lines, the profiles do not match exactly. This could be due to  a
number of different factors: a greater spatial extent for the \HISA\ than for the $^{12}$CO,
different beamsizes of
the observations, or errors in the background subtraction of the \HISA\ features.
  Positions 7 though 9 are not shown due to the lack of 
molecular emission at those positions.

\subsection{Molecular Gas Column Densities}
\label{models}

There are a number of techniques for solving the equations of radiative  
transfer and detailed balance in molecular clouds.  However, to determine the  
physical parameters of the molecular components of the globule and the   
complex, we used a Large Velocity Gradient (LVG; i.e. Goldsmith et  
al. 1983)  model which offers a rudimentary approach to  
determining the bulk properties of a region based on the integrated  
intensities of the observed lines. These models employed  the collision rates of Flower \& Launay (1985) and 
Shinke et al. (1985).  

In this section, we describe the method by which we
determined the molecular gas column densities from our CO observations.  
The results of this analysis will be used in 
Section \ref{discuss} to constrain the physical properties of the \HISA\ clouds, as well
as the atomic gas fraction.

\subsubsection{The Globule}
\label{glob}

To determine the molecular gas content in the \HISA\ globule itself, we are  
interested in $^{12}$CO emission at V$_{LSR}$ = -41 km s$^{-1}$. However, since there is
no detectable signal at this velocity we set an  upper limit to the  
$^{12}$CO column density by using the 1$\sigma$ rms noise of the $^{12}$CO \trans{2}{1} transition ($T_{mb}$ =  0.06 K).
In lieu of any actual CO detection in the globule, we used a range of parameters chosen to cover the likely range of physical conditions.  Thus,
for the globule, we used kinetic temperatures
from 8 K to 50 K and fixed the density at $10^2$, $10^3$, and $10^4$ cm$^{-3}$ to check for differences between sub-thermal and LTE excitation of the $^{12}$CO lines.   For each temperature-density combination, we used our LVG code to calculate the $^{12}$CO \trans{2}{1} brightness for a series of 50 different column densities.  The column densities ranged from $5\times10^{14} - 5\times10^{17}$ cm$^{-2}$ and were logarithmically-spaced.  Thus, by comparing our 1$\sigma$ noise limit to the LVG models, we were able to set an upper limit to the   $^{12}$CO column density for each temperature-density combination.
The maximum $^{12}$CO column density, corresponding to the lowest temperature-density combination (8K and 100 cm$^{-3}$ respectively), is $9.5\times10^{15}$ cm$^{-2}$.

At such small column densities, $^{12}$CO self-shielding is very inefficient and  
the H$_2$/$^{12}$CO abundance ratios can be strongly affected. However, using the 
photodissociation region (PDR) models of van Dishoeck  
\& Black (1988), and assuming that the strength of the UV field is that of the average interstellar radiation field (i.e. G$_o$ = 1), we can estimate the total H$_2$ column density in the globule.  For our 
maximum $^{12}$CO column density of $9.5\times10^{15}$ cm$^{-2}$, the van Dishoeck  
\& Black curves provide an H$_2$ column density of 
N(H$_2$) $< 1.1 \times 10^{21}$ cm$^{-2}$.

\subsubsection{The Complex}
\label{complex}

 For the complex, we only modeled the spectrally smoothed data since they are most directly comparable to the \HISA\ spectra.  In our models we used kinetic temperatures of 12 K to 50 K (at temperatures lower than 12K we were unable to find a LVG fits to our $^{12}$CO data).  At each
temperature
we created a $50\times 50$ logarithmically-spaced grid of LVG models in density-column density parameter space. 
The densities ranged from 
$10^2 - 5\times10^4 $ cm$^{-3}$, and 
the $^{12}$CO column densities  from $5\times10^{15} - 5\times10^{18}$ cm$^{-2}$.  
 The observed line intensities were fit to the grid of LVG models using a 
$\chi^2$ minimization routine to find the density$-$column density combination that best fit the 
observations at each temperature.
Again, comparing our $^{12}$CO column densities to the models of van Dishoeck \& Black  
(1988), we  also determined the H$_2$ column densities.   The results of this LVG analysis will be presented in Section \ref{subsec:A+M}.

To test our assumption that the multiple velocity components seen in  
the complex were due to separate clouds rather than self-absorption in  
a single cloud, we also ran an extensive series of  Accelerated Monte Carlo 
(AMC; Hogerheijde \& van der  
Tak, 2000) models. AMC models have the advantages of  
producing model spectra based on the input physical parameters of the  
cloud (i.e. temperature, density and velocity gradient), and a greater  
available range of parameter space.  While LVG models incorporate the  
simplifying assumption that the concentric model shells are radiatively  
decoupled, this is not the case for the more robust AMC models.   

We first  
attempted to model the cloud as a single entity with a variety of   
density, temperature, abundance, and velocity gradients.   Some models,  
in which the abundance ratio was varied with a power law on the order  
of 0.5 came close to matching the $^{12}$CO spectra, but no single  
model (for any position) was able to match the $^{12}$CO and $^{13}$CO  
line profiles and ratios, despite attempts at varying the  
$^{12}$CO/$^{13}$CO abundance ratio.  In addition, none of these models were  
capable of matching the shift in the red-blue asymmetry discussed  
previously.  In all, we ran several thousand models.  The total  
parameter space covered is shown in Table \ref{tab:AMCspace}.
Column 1 
shows the parameters that were varied and column 2 shows the range of central values used for each  parameter.  Models were run as powerlaw functions of the form 
$x(r) = x_o\left(\frac{r_o}{r}\right)^{\alpha}$ where  
$x_o$ is the value of the parameter (abundance, temperature, density) at some characteristic radius ($r_o$), $\alpha$ is the power law index , and  $r$ is
the radius.
$\alpha$ 
was allowed to vary between -2 and +2, except for variations in the density for which $\alpha > 0$ would be
unphysical.
 We  also tried modeling each individual position as a separate and distinct  
``cloudlet'', with its own unique temperature, density and velocity gradients.   
Again, none of the  models were able to reproduce the spectra across  
the entire complex. Therefore, we are reasonably confident that 
the multiple velocity components seen in the spectra are due to separate line of sight  
clouds.

\section{Discussion}
\label{discuss}

\subsection{Atomic and Molecular Gas Components in \HISA\ Clouds}
\label{subsec:A+M}
 
 Gibson et al. (2000) presented a preliminary view of HISA features seen in the   CGPS, finding a number of small scale HISA features that had not previously been seen in single-dish surveys.  In  more recent papers (Gibson et al. 2005a \& 2005b) present a much more detailed view of HISA features through a re-analysis of the CGPS data, using automated algorithms to locate HISA features in the HI data cube instead of the ``by-eye'' identification used in the earlier paper.
In addition, Gibson et al (2005b)  contains a  revision of the \HISA\ data presented in G2000.  In the earlier paper the \HISA\ amplitudes were in error; the correct \HISA\ amplitudes are lower than those listed in G2000.
In this section we use these improved \HISA\ observations and the technique presented in Gibson et al. (2000) 
to determine the physical  
properties of \HISA\  gas.  

The technique presented by G2000 allows one to constrain the 
physical properties of the \HISA\ gas in the complex and the globule by 
placing limits on the optical depth and spin temperature of the \HISA\ features.
 The optical  
depth of the \HISA\ can be determined through:

\begin{equation}
\tau_{_{HISA}} = \frac{CN_{tot}f}{T_K\Delta v}
\label{eqn:tau}
\end{equation}

\noindent Where $C$ is   combination of constants for the  
{\sc HI} spin-flip transition 
= 5.2$\times10^{-19}$ (Dickey \& Lockman 1990), $\Delta v$ is the line  
width, $T_K$ is the kinetic temperature of the gas which, under the  
assumptions of G2000 should be the same as $T_s$, the {\sc Hi} spin  
temperature, and $N_{tot}$ is the {\it total} column density (i.e.  
$N_{atomic} + N_{molecular}$).

This analysis was able to find families of solutions based upon
allowed ranges of the unknown atomic gas fraction ($f = $N(HI)/[N(HI) + N(H$_2$)]) and $p$, 
the fraction of HI emission originating behind the HISA cloud.  Figure 6 of G2000 shows
their results.  Even if we make the reasonable assumption that $p$ is close to unity, indicating that
most of the cold HI is in the foreground as would be needed for strong HI self-absorption, the unknown atomic gas fraction still limits our ability to determine the amount and temperature of the \HISA\ gas.  For example, Figure 6 of G2000 shows that in the complex, if the \HISA\ is warm ($T_k > 40$ K) then $f \sim 1$ and the \HISA\ opacity ($\tau_{HISA}$) is approximately 1 to 2, whereas if the \HISA\ is cold ($T_k < 10$K) then $f < 0.1$ and $\tau_{HISA}$ is constrained to$\sim 0.4 - 0.5$.

With our LVG analysis of the $^{12}$CO observations, we are able to better
constrain the \HISA\ properties ($f$, $T$, and $\tau$) in the globule and complex, by finding the overlap between the original parameter space of G2000 and the parameter space defined by our $^{12}$CO observations.
For the \HISA\ gas, this was done according to the procedure detailed in G2000 but using the corrected
data as given in Gibson et al. 2005b.  
For our $^{12}$CO observations, we took the derived H$_2$ column density for each position (Section \ref{models})
and, using the above equation, calculated $\tau_{HISA}$ for a range of atomic gas fractions ($f$) from
0.01 to 1.  

\begin{figure}
\begin{center}
\plotone{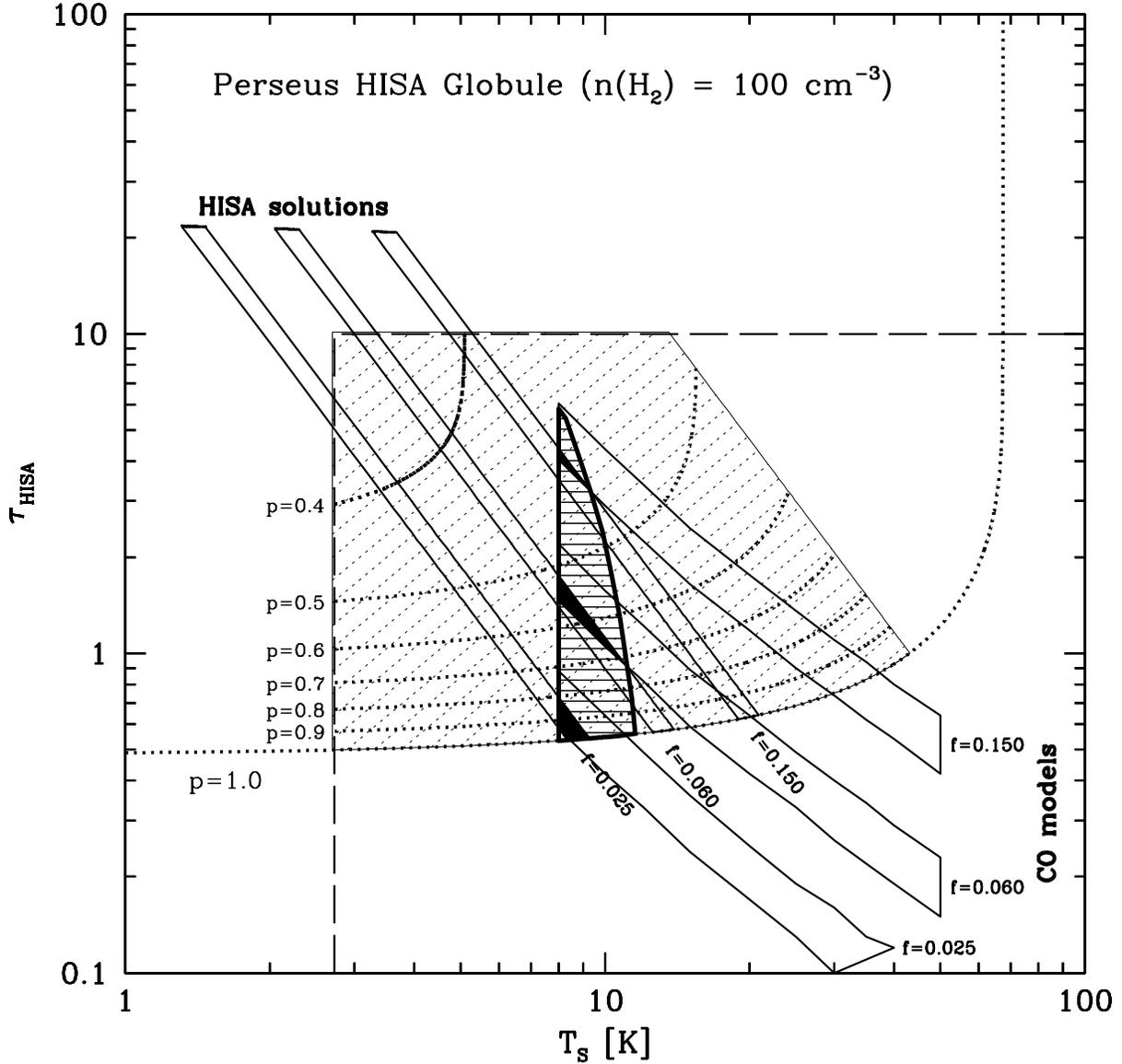}
\caption{Physical parameters of the \HISA\ gas in the globule as derived by atomic and molecular  
observations.  The lightly shaded  area shows the parameter space determined from the original \HISA\ analysis presented in G2000.  
The dotted curves show the values of $p$, the thin, tilted strips (starting at the upper-left corner of the plot) shows $\tau_{HISA}$ vs $T_{spin}$ as determined 
from the Gibson et al. (2000) analysis of the \HISA\ features for a range of assumed atomic gas fraction ($f$) , and the wider, tilted strips (starting in the bottom-right corner of the plot) shows 
 $\tau_{HISA}$ vs $T_{spin}$ as determined from our $^{12}$CO analysis for a range of atomic gas fractions.  The bold
``sharkfin"-shaped region shows the union of the \HISA\ and $^{12}$CO  solutions for all values of $f$, where each individual intersection is for a particular $f$ only. 
}
\label{fig:hisa}
\end{center}
\end{figure}

Figure \ref{fig:hisa} shows an example for the Globule in which we plot $\tau_{HISA}$ vs $T_{spin}$.  For this figure we have assumed that n(H$_2$) $= 10^2$ cm$^{-3}$ in our LVG calculation.  
The dotted curves show the values of $p$, the thin, tilted strips (starting at the upper-left corner of the plot) show $\tau_{HISA}$ vs $T_{spin}$ as determined 
from the Gibson et al. (2000) analysis of the \HISA\ features for a range of assumed atomic gas fraction ($f$) , and the wider, tilted strips (starting in the bottom-right corner of the plot) show
 $\tau_{HISA}$ vs $T_{spin}$ as determined from our $^{12}$CO analysis for a range of atomic gas fractions.  The bold
``sharkfin"-shaped region shows the union of the \HISA\ and $^{12}$CO  solutions for all values of $f$, where each individual intersection is for a particular $f$ only.  As can be seen from Figure \ref{fig:hisa}, the inclusion of the $^{12}$CO data significantly constrains the 
allowed range of physical conditions in the \HISA\ gas.  The ``sharkfin" corresponds to a solution of 8 K $< T_{spin} < 11$ K and $0.5 < \tau_{HISA} < 5.8$.  If we
make the further assumption that $p > 0.8 $ (i.e. that most of the Hi emission is in the background as would be required to produce strong \HISA\ features), then $0.5 < \tau_{HISA} < 0.8$.  

In addition, since the intersection of \HISA\ and $^{12}$CO solution also corresponds to an intersection
of $f$, we can constrain the atomic gas fraction, and find that $f = 0.02 - 0.06$.  A similar analysis was done for assumed densities of $10^2$ and $10^3$ cm$^{-3}$ in the globule, and for each position in the complex.  The results are given in Table \ref{tab:hisa}.
Note that no solutions were found for the Complex - Positions 1 \& 2.  This is probably due to the presence of the strong second spectral feature.  Since the LVG analysis was done on the spectrally smoothed data (to match the spectral resolution of the \HISA\ observations) the two line-of-sight clouds are treated as a single component (see Figure \ref{fig:smoothCO}).  The atomic gas fraction in Positions 3 through 6 were
found to be $f = 0.02 - 0.1$.
The combination of the \HISA\ and molecular observations also allow us to better constrain the molecular gas density and column density since we now have a better constraint on the kinetic temperature. The results are given in Table \ref{tab:molgas}.

Given the small atomic gas fractions, the complex and the globule seem to be  predominantly 
molecular in composition.  While this seems obvious for the central positions in the complex where
the $^{12}$CO lines are relatively strong, it is less obvious for Position 6, where the $^{12}$CO lines are weak.  Nevertheless  our analysis suggests that $> $ 90\% of the gas is molecular.  It is even more puzzling in the globule where the $^{12}$CO lines are undetectable and yet our limits on the $^{12}$CO column density suggest that up to 95\% to 98\% of the gas could be molecular.  Thus it is possible that the globule and the edge of the complex are regions where the gas has a large molecular component that is not well-traced by $^{12}$CO.  There are, however, possible alternative explanations.  Gibson et al (2000) derived the spin temperature of the \HISA\ features via the equation

\begin{equation}
T_s = \sqrt{\frac{\langle P \rangle f_nC\Delta s}{k \tau_o \Delta V}}
\label{eqn:ts}
\end{equation}

Two main
assumptions that go into producing equation  (2) are: a) the volume density is equal to the column density divided by the path length through the cloud ($\Delta s$) and, b) the volume density is related to the spin temperature through the ideal gas law. Therefore, if either the gas pressure $\langle P \rangle$ is below the canonical value of $P/k = 4000$ K cm$^{-3}$, or the \HISA\ features are thinner along the line-of-sight than assumed by G2000, then the  atomic gas fraction would have to be larger than predicted to produce the same spin temperature.  However, to produce gas which is primarily atomic (i.e. $f \sim 1$) would require either a large drop in gas pressure or significantly foreshortened clouds.
While there is no evidence to support the hypothesis of significantly lower than average gas pressures, there is evidence to support the notion that observationally identified clouds may be preferentially {\it elongated} along the line of sight rather than foreshortened (e.g. Heiles 1997) due to observational selection effects.

We can estimate the total mass contained in the 45$''$ beam of our FCRAO $^{12}$CO $J = 1\rightarrow 2$ observations from the solutions given above and using a distance to Perseus of 2 kpc.  The upper limit for the total
gas mass in the globule (M(HI) + M(H$_2$)) is $1.3 - 3 M_{\odot}$ for the given range of temperatures, H$_2$ column densities, and atomic gas fractions.  This range is between 7 and 60 times lower than the
Jeans masses calculated for the globule for the given ranges of temperature and density.  A similar calculation for the complex finds that each 45$''$
beam  contains $3 - 4 M_{\odot}$ of material.  This is again 7 to 60 times lower than the Jeans mass implying that neither the globule nor the individual positions in the complex are gravitationally bound.

\subsection{The Nature of the \HISA\ Clouds}
\label{nature}

So what are these \HISA\ clouds?  
They appear to be non-gravitationally bound regions of cold, primarily molecular, gas that is not detected in $^{12}$CO.
Such regions are not unusual in high-latitude cirrus clouds and have  been traced via excess IR emission (e.g. Heiles, Reach, \& Koo 1988; Reach, Koo \& Heiles 1994).  However, most of the cirrus
clouds in which $^{12}$CO does not seem to trace the total amount of molecular gas are relatively warm 
(T$_K > 30$K), low density (n(H$_2$) $< 100$ cm$^{-3}$) regions in which the HI does not appear to be
self absorbed.  In contrast, the temperatures (10 K - 25 K) and densities (n(H$_2$) $\sim 100 - 1200$ cm$^{-3}$) in our \HISA\ clouds are similar to those seen in molecular clouds.  Thus,  the \HISA\ gas has temperatures and densities that bridge the gap between the ambient atomic ISM and the colder,
denser molecular medium.

Could these \HISA\ clouds be sites where atomic gas is condensing into the molecular phase
required for star formation?  Although critical to the evolution of matter in
the Galaxy, molecular condensation is poorly understood.  
Using a model that assumes that molecular clouds form from atomic gas after the passage of shock waves, Bergin et al. (2004) find that the 
molecular cloud formation timescale is not controlled by the formation rate of H$_2$ on grains but, rather, by the shielding of molecules from the UV radiation.  While the  H$_2$ can self-shield quite efficiently, $^{12}$CO formation requires shielding of the interstellar radiation by dust grains. If the total A$_V$ is greater than $\sim 0.7$ then there is enough material present to effectively shield the $^{12}$CO.  

Using the
60 and 100 $\mu$m IRAS HIRES data, we have estimated the dust temperature and the
amount of visual extinction in the globule and at each of our observed positions in the complex.  Following the procedure outlined in Wood et al. (1994), we calculated the dust temperature from the ratio of the 60 and 100 $\mu$m fluxes, assuming that the dust is optically thin, that the 60 and 100 $\mu$m beam solid angles are roughly equivalent, and that the dust emissivity spectral index is 1.5.  We find that the dust temperatures are all $< 30$K.   If the dust is optically thin, knowing the temperature allows us to determine the 100 $\mu$m dust opacity  from the ratio of the observed 100 $\mu$m flux to the Planck function.  The dust visual extinction was then calculated from the
relationship between A$_V$ and the 100 $\mu$m dust opacity provided by Wood et al. (1994), which is a functional fit to the data given in Jarrett et al. (1989).   In the complex, the dust visual extinction ranges from
A$_V$ = 3.8 at Position 1 to A$_V$ = 2.2 at Position 6.  For Positions 7 through 9, A$_V$ drops below 2, as it does in the globule where A$_V$ = 1.2.  Note that these are the visual extinctions through the {\it entire} cloud.  The edge-to-center visual extinctions, which are directly comparable to the Bergin et al. (2004) models,  are half these values.  Thus, these results are consistent with the scenario given by Bergin et al (2004)  in which the absence of detectable $^{12}$CO in the globule, and in Positions 7 - 9 of the complex, is due to the limited UV shielding provided by the dust in these regions.  
However, there could
still be a considerable amount of molecular gas in Positions 7 - 9 and in the globule, since
H$_2$ can form at considerably earlier times and lower column densities than $^{12}$CO.

Since A$_V$ in the globule and at the edge of the complex is close to the critical value needed to
shield $^{12}$CO from the interstellar radiation field, it is possible that these regions are in the process of
forming $^{12}$CO.    Thus, over time, $^{12}$CO lines may eventually become observable as the $^{12}$CO abundance 
continues to increase.
Bergin et al. (2004) predict the observed intensities of various transitions as a function of time (as clouds evolve from atomic to molecular).  In their model, a cloud with a $^{12}$CO $J = 2\rightarrow 1$ line strength less that 0.06 K (our 1$\sigma$ detection limit) would  be less than $10^7$ years old.  This ``age''
is consistent with the minimum transit time between spiral arms in the outer Galaxy ($\sim 10^7$ years; Heyer \& Tereby 1998).

An alternative scenario is that instead of being molecular clouds in the process of formation, the \HISA\ clouds could represent transient events, or even the dispersal of molecular clouds into the atomic medium.  While we cannot rule out these possibilities, the formation scenario seems more likely since the Complex and the Globule both seem to be correlated with a region slightly downstream of the spiral shock wave in the Perseus arm (Gibson et al 2005; Roberts 1972) where the gas is densest.  This is precisely the type of region  where we would expect H$_2$ to be condensing from HI.

\section{Conclusions}
\label{sec:conclusions}

Using $^{12}$CO $J = 2\rightarrow1$ observations from the CSO, and 
$^{12}$CO  and $^{13}$CO $J = 1\rightarrow0$ observations from the FCRAO, we have determined the molecular gas content in two regions of HI Self-Absorption (\HISA) in Perseus.  In the globule we observed a small 8-point map at the CSO to match 
the 45$''$ resolution, single-point $J = 1\rightarrow0$ observation taken with the FCRAO.  In the 
complex we observed a nine point strip  from the center of the cloud  to the edge of the cloud in increasing galactic latitude.  No $^{12}$CO $J = 2\rightarrow1$ emission was detected
in the globule  to a 1$\sigma$ rms limit of 0.06 K.  $^{12}$CO was detected in Positions 1 through 7 in the complex but fell below the 
1$\sigma$ rms noise limit in Position 8 and 9.  Positions 1 \& 2 were found to contain two
spectral features which we determined to be due to separate line-of-sight clouds rather than to $^{12}$CO self absorption.

Using both Large Velocity Gradient  and Monte Carlo radiative transfer codes, we were able to
determine the molecular gas content in the globule and complex.
In the globule N($^{12}$CO)  $< 6.0 \times 10^{15}$ cm$^{-2}$, implying
that  N(H$_2$) $< 9.9 \times 10^{20}$ cm$^{-2}$.
In the complex we found that the H$_2$ column densities ranged from  $1.2 - 2.2 \times 10^{21}$ cm$^{-2}$.

By comparing the \HISA\ and $^{12}$CO observations we are able to constrain the physical conditions 
and atomic gas fraction ($f$) of these two regions.  In the globule,  8 K $< T_{spin} < 22$ K and $0.02 < f < 0.2$ depending on whether the (unknown) gas density is $10^2$, $10^3$, or $10^4$ cm$^{-3}$.  In the complex, 12 K $< T_{spin} < 24$ K, $0.02 < f < 0.05$, and the gas density is constrained ($100 < n < 1200$ cm$^{-3}$.  These results imply that the gas in the \HISA\ clouds is colder and denser than that
usually associated with the atomic ISM and, indeed, is similar to that seen in molecular clouds.
The small atomic gas fractions also imply that there is a significant molecular component in \HISA\ clouds, even when little or no $^{12}$CO is detected. The level of $^{12}$CO detected and the visual extinction due to dust is consistent with the idea that these \HISA\ clouds are undergoing a transition from the atomic to molecular phase.

\acknowledgments

The authors would like to thank Floris van der Tak for his help
with  the Monte Carlo Models, and the Natural Sciences and Engineering Research  
Council of Canada for their financial support.  The Five College Radio
Astronomy Observatory is operated with the permission of
the Metropolitan District Commission, Commonwealth of
Massachusetts, and with the support of the National Science
Foundation under grant AST 01-00793.  The CSO is funded 
under a grant from the National Science Foundation.

\clearpage


\begin{table}
\begin{center}
\caption{$^{12}$CO Gaussian Fits in the Globule}
\begin{tabular}{cccccccc}
\hline \hline
       & & \multicolumn{3}{c}{\trans{1}{0} transition} &  \multicolumn{3}{c}{\trans{2}{1} transition} \\
      \hline
 $\Delta \alpha$  & $\Delta \delta$  & $T_{mb}$ & V$_{LSR}$ &  FWHM & $T_{mb}$ & V$_{LSR}$ &  FWHM \\
($''$)  &  ($''$) &  (K)  & (km s$^{-1}$) & (km s$^{-1}$) &  (K)  & (km s$^{-1}$) & (km s$^{-1}$) \\

\hline
\multicolumn{8}{c}{32$''$ Resolution}\\
\hline
0 	& 0     & - & - & - & 0.33    & -45.1 & 1.13 \\
0       & -64   & - & - & - & $<$0.14 & - & - \\
 +32    & -32   & - & - & - & $<$0.20 & - & - \\
 +32    & 0     & - & - & - & $<$0.13 & - & - \\
 0      & +32   & - & - & - & $<$0.07 & - & - \\
 -32    & 0     & - & - & - & $<$0.13 & - & - \\
 -32    & -32   & - & - & - & $<$0.12 & - & - \\
 0      & -32   & - & - & - & 0.52    & -45.1 &1.5  \\
 0      & -16   & - & - & - & 0.54    & -45.0 & 1.21  \\
 \hline
 \hline
\multicolumn{8}{c}{Convolved to 45$''$ Resolution}\\
\hline
0 & 0 & $<$0.58 & - & - & 0.35  &  -45.0  &  1.2 \\
 \hline \hline

\end{tabular}
\label{tab:glob}
\end{center}
\end{table}

\clearpage

\begin{table}
\begin{center}
\caption{$^{12}$CO Gaussian Fits in the Complex}
\begin{tabular}{lcccccccc}
\hline \hline
 &      & & \multicolumn{3}{c}{\trans{1}{0} transition} &  \multicolumn{3}{c}{\trans{2}{1} transition} \\
      \hline
 Position & $\Delta \alpha$  & $\Delta \delta$  & $T_{mb}$ & V$_{LSR}$ &  FWHM & $T_{mb}$ & V$_{LSR}$ &  FWHM \\
 & ($''$)  &  ($''$) &  (K)  & (km s$^{-1}$) & (km s$^{-1}$) &  (K)  & (km s$^{-1}$) & (km s$^{-1}$) \\

\hline
& & & \multicolumn{3}{c}{45$''$ Resolution} & \multicolumn{3}{c}{32$''$ Resolution}\\
\hline
1 & 0  & 0                    & 2.7 &   -41.4 &  1.2 &    2.4 &    -41.7 &  1.3  \\
 ...   & ...    &  ...        & 9.5 &   -39.8 & 1.0  &   6.4 &   -39.7 & 1.5 \\
2 & +12.9$''$   & +22.5$''$   & 7.1 &   -41.4 &  1.4 &    3.1 &    -41.6 &  1.5  \\
 ...   &  ...   &  ...        & 8.5 &   -39.8 & 1.3  &   5.7 &   -39.8 & 1.4 \\
3 & +25.8$''$   & +45.0$''$   & 8.3 &   -40.8 &  2.6 &    6.3 &    -40.7 &  2.3 \\
4 & +38.8$''$   & +67.8$''$   & 8.3 &   -41.0 &  2.2 &    5.8 &    -41.0 &  2.2  \\
5 & +51.6$''$   & +90.3$''$   & 7.6 &   -41.0 &  2.0 &    4.6 &    -41.1 &  2.0  \\
6 & +64.5$''$   & +112.8$''$  & 3.6 &   -41.1 &  2.4 &    1.3 &    -41.1 &  2.1 \\
7 & +77.4$''$   & +135.3$''$  & $<$0.22 &   - & - & $<$0.08 & - & - \\
8 & +90.5$''$   & +158.1$''$  & $<$0.19 &   - & - & $<$0.07& - & - \\
9 & +103.4$''$  & +180.6$''$  & $<$0.25 &   - & - & $<$0.06 & - & - \\ 
 \hline
 \hline

\end{tabular}
\label{tab:com}
\end{center}
\end{table}

\clearpage

\begin{table}
\caption{AMC Model Parameters for the Complex.}
\begin{center}
\begin{tabular}{lc}
\hline\hline
Parameter   & Range    \\

\hline
Kinetic Temperature   & 10 - 70  K   \\
Volume Density     & 100 - 6000  cm$^{-3}$ \\
$^{12}$CO/H$_2$ Abundance Ratio &10$^{-4}$ - 10$^{-7}$   \\
Radial Velocity & 0 - 3   km s$^{-1}$      \\
Turbulent Linewidth& 0.05 - 1  km s$^{-1}$   \\
Cloud Radius       &10$^{15}$ - 10$^{19}$ cm \\
\hline
\end{tabular}
\label{tab:AMCspace}
\end{center}
\end{table}

\clearpage

\begin{table}
\caption{Range of Physical Conditions in HISA Clouds (assuming $p > 0.8$)}
\begin{center}
\begin{tabular}{lcccccccc}
\hline\hline
Position & $T_{spin}$(min) & $T_{spin}$(max) & $\tau_{HISA}$(min) & $\tau_{HISA}$(max)     \\
\hline
Complex - Position 1 & \multicolumn{4}{c}{no solution} \\
Complex - Position 2 & \multicolumn{4}{c}{no solution} \\
Complex - Position 3 & 14.5 & 23.7 & 0.4 & 0.5 \\
Complex - Position 4 & 14.5 & 22.4 & 0.3 & 0.6 \\
Complex - Position 5 & 13.9 & 20.0 & 0.4 & 0.6 \\
Complex - Position 6 & 12.0 & 18.0 & 0.4 & 0.5 \\
Globule (n = $10^2$ cm$^{-3}$)  & 8.0 & 11.0 & 0.5 & 0.8 \\
Globule (n = $10^3$ cm$^{-3}$)  & 10.0 & 20.2 & 0.6 & 0.9 \\
Globule (n = $10^4$ cm$^{-3}$)  & 11.0 & 22.1 & 0.6 & 0.9 \\
\hline
\end{tabular}
\label{tab:hisa}
\end{center}
\end{table}

\clearpage

\begin{table}
\footnotesize
\caption{Molecular Gas Properties}
\begin{center}
\begin{tabular}{lccccccccc}
  \hline
  \hline
        & \multicolumn{2}{c}{J = $1\rightarrow 0$} &  
\multicolumn{2}{c}{J = $2\rightarrow 1$}&\multicolumn{2}{c}{}\\
Position   &T$_{mb}$  & $\Delta V_{FWHM}$ &  T$_{mb}$  & $\Delta V_{FWHM}$  &  
n(H$_2$) & N($^{12}$CO) &  N(H$_2$) & T$_K$ \\
 & K & km s$^{-1}$   & K &  km s$^{-1}$ & cm$^{-3}$ & cm$^{-2}$ & K & cm$^{-2}$  \\

\hline

Complex - Position 1  &   8.6 $^a$  &   1.3 & 5.0 &  2.6 & 120 & 3.2$\times10^{17}$ & $2.2\times10^{21}$ & 20 $^b$ \\
Complex - Position 2  &   8.0 $^a$ &   2.9 & 4.6 &  2.7 & 370 & 1.2$\times10^{17}$  & $1.8\times10^{21}$ & 20 $^b$\\
Complex - Position 3  &   8.2 $^a$ &   2.6 & 6.1 &  2.4 & 2220 & 3.2$\times10^{16}$ & $1.3\times10^{21}$ & 20  \\
Complex - Position 4  &   8.0  $^a$ &   2.3 & 5.7 &  2.3 & 1840 & 2.9$\times10^{16}$ & $1.2\times10^{21}$ & 20  \\
Complex - Position 5  &   7.3  $^a$ &   2.1 & 4.5 &  2.1 & 1150 & 4.3$\times10^{16}$ & $1.3\times10^{21}$ & 15 \\
Complex - Position 6  &   3.6  $^a$ &   2.5 & 1.3 &  2.1 & 140 & 5.7$\times10^{16}$ & $1.5\times10^{21}$ & 15  \\
Globule                         &   $< 0.58$  &   - & $< 0.06$ &  -  & 100 & $< 6.0\times10^{15}$ & $9.9\times10^{20}$ & 10 \\
Globule                         &   $< 0.58$  &   - & $< 0.06$ &  -  & 10$^3$ & $< 4.0\times10^{14}$ & $5.3\times10^{20}$ & 15  \\
Globule                         &   $< 0.58$  &   - & $< 0.06$ &  -  & 10$^4$ & $< 1.0\times10^{14}$  & $4.2\times10^{20}$  & 15\\

\hline
\end{tabular}
\label{tab:molgas} 
\end{center}
$^a$ - Spectral line properties from the spectrally smoothed data. \\
$^b$ - Assumed temperature based on \HISA-CO analysis in Positions 3 - 6 (Table \ref{tab:hisa}).

\end{table}

\clearpage

\clearpage

\clearpage

\clearpage

\clearpage

\end{document}